# The Blandford-Znajek mechanism with Vacuum Breakdown as a Gamma-ray Burst Progenitor

Guido Barbiellini[*] and Francesco Longo[†]

[*]Department of Physics University of Trieste and INFN, Trieste
[†]Department of Physics University of Ferrara and INFN, Ferrara

**Abstract.**
The energetics of the long duration GRB phenomenum is compared with the BZ mechanism. A rough estimate of the energy extracted from a rotating Black Hole with the Blandford-Znajek mechanism is evaluated with a very simple assumption: an inelastic collision between the rotating BH and an accreting torus. The GRB energetics requires an high magnetic field that breaks down the vacuum around the BH and gives origin to a e$^{\pm}$ fireball.

## PHENOMENOLOGICAL OVERVIEW

Gamma-ray Bursts (GRBs) until a few years ago were largely devoid of any observable counterpart at any other wavelengths. However, a dramatic development in the last several years has been the measurement and localization of fading X-ray signals from some GRBs, lasting typically for days and making possible the optical and radio detection of afterglows, which mark the location of the GRB event. These afterglows in turn enabled the measurement of redshift distances, the identification of host galaxies, and the confirmation that GRB were at cosmological distances (for a recent brief review see [1]). The temporal decay of the emission in different frequencies for several GRBs has been interpreted according to the fireball model and suggested jet beaming with opening angle $\theta \sim 4°$ [2].

Another important discovery made in the last year is the presence of iron lines in the X-ray spectrum of GRBs (for example [3, 4, 5]). This provides a powerful tool to understand the nature and the environment of GRB primary sources [6, 7]. The presence of strong iron lines implies a rich environment located very close to the GRB and it may be an argument in favour of massive-star progenitor models of GRB [8, 9, 10, 11].

The presence of an iron cloud is in favour of the interpretation of GRBs as a second step of the residual of the primary explosion(e.g.[11]). In this interpretation the primary explosion leaves over a compact object that could be a rotating black hole, at the center of the environment consisting of ejecta. In this scenario, it is plausible the hypothesis of energy extraction from a rotating BH (compact object left over from the primary explosion), through the Blandford-Znajek mechanism [12], where the external magnetic field can be supplied by an Fe torus circulating around the BH at a distance R (of the order of R$_s$).

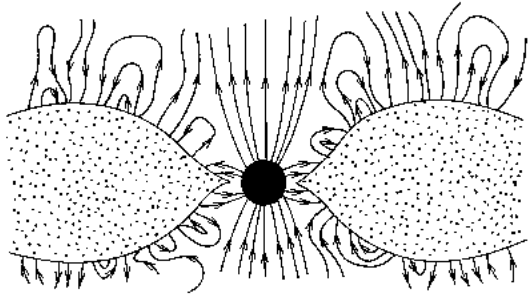

**FIGURE 1.** Possible configuration of the magnetic field close to the BH surrounded by an accretion disk (from [13])

## GAMMA-RAY BURST PROGENITOR

At cosmological distances the observed GRB fluxes imply energies of order of up to a solar rest-mass ($\sim 10^{54}$ erg), and from causality these must arise in regions whose size if of the order of kilometers in a time scale of the order of seconds. This implies that an $e^{\pm}, \gamma$ fireball must form, which would expand relativistically. The short life time of the GRB and the large distance suggest as the GRB source a compact system undergoing a very energetic transition with high an efficient mechanism for conversion of energy into $\gamma$-ray emission. Accreting black holes are so good candidates for being source of GRBs.

Blandford and Znajek have proposed an interaction between a rotating BH and an accretion disk to explain the energetics of Active Galactic Nuclei[12]. The same mechanism could be a good candidate for GRB engines[10, 14]. In the BZ mechanism the magnetic field of the accretion disk acts as a brake on the BH, and the energy output is mainly due to the loss of rotational energy. The rotational energy for a maximally rotating BH, with the rotation parameter $\tilde{a} = \frac{Jc}{M^2 G} = 1$, is $0.29\ Mc^2$. Even with the optimal efficiency the energy availaible for the BZ mechanism is [14]:

$$E_{BZ} \simeq 0.3 \cdot 10^{54} \left(\frac{M}{M_\odot}\right) \text{erg}$$

In the following considerations it we assume that a dissipative interaction is at work between the BH and the torus surrounding it. The dissipative interaction is due to an internal torque. If the short interaction is treated as an inelastic shock it is possible to apply the angular momentum conservation law ( in its classical mechanics approximation ).

$$I_{\text{BH}}\Omega_{\text{BH}} + I_{\text{disk}}\Omega_{\text{disk}} = I\Omega$$

where $I_{\text{BH}}$ and $I_{\text{disk}}$ are the momenta of inertia of the BH and the disk, $\Omega_{\text{BH}}$ and $\Omega_{\text{disk}}$ are the angular velocities. $I$ and $\Omega$ are the same quantities for the resulting BH. In this approximation the loss of kinetic energy is

$$\Delta E = \frac{1}{2}\frac{I_{BH}I_{disk}}{I_{BH}+I_{disk}}(\Omega_{BH}-\Omega_{disk})^2$$

Let us consider as a typical configuration a BH with a mass of 10 $M_\odot$ with a rotating disk of 0.1 $M_\odot$ at the last stable orbit $R_l \sim 3R_s$:

$$I_{\text{disk}} = M_{\text{disk}} \cdot 9R_s^2 \qquad I_{\text{BH}} = \tfrac{2}{5} M_{\text{BH}} R_s^2$$
$$I_{\text{disk}} < I_{\text{BH}} \qquad \Delta E = \tfrac{1}{2} I_{\text{disk}} (\Omega_{\text{BH}} - \Omega_{\text{disk}})^2$$

Assuming that $\Omega_{\text{disk}} < \Omega_{\text{BH}}$, the energy availaible for the BZ process to power the GRB is:

$$\Delta E = \frac{1}{2} I_{\text{disk}} \Omega_{\text{BH}}^2 \sim 8 \cdot 10^{53} \left( \frac{M_{\text{disk}}}{0.1 M_\odot} \right) \text{erg}$$

where we have used $I_{disk} = 9 M_{disk} R_s^2$ and $\Omega_{BH} \sim \frac{c}{R_s}$.

## VACUUM BREAKDOWN

If the environment is sufficiently clean, the GRB fireball could be generated by the vacuum breakdown[15] in the volume close to the polar cap of the BH. The magnetic field required to explain the high luminosity of GRB generates an electric field that could break down the vacuum.

The luminosity of the BZ mechanism is [10]:

$$L_{BZ} \sim 10^{51} \left( \frac{B}{10^{15} \text{G}} \right)^2 \left( \frac{M_{\text{BH}}}{10 M_\odot} \right)^2 \text{erg s}^{-1}$$

The BZ mechanism generates a voltage drop of the order [13]:

$$\Delta V = 10^{23} \cdot \left[ \frac{M}{10 M_\odot} \right] \left[ \frac{B}{10^{15} G} \right] \text{V}$$

Assuming a mass of the BH of the order of 10 $M_\odot$, to this voltage drop corresponds an electric field in the proximity of the BH:

$$E = \frac{\Delta V}{2\pi R} = 10^{23} \frac{\frac{B}{10^{15} G}}{6\pi \cdot 10^4} \frac{\text{V}}{\text{m}}$$

The energy acquired by an electric charge $e$ over the reduced Compton electron wavelenght $\bar{\lambda} = \frac{\hbar}{m_e c}$ under this electric field can exceed $2 \cdot m_e c^2$ if:

$$eE\bar{\lambda} > 1.6 \cdot 10^{-6} \text{erg}$$

From this relation it is possible to derive an estimate for the magnetic field:

$$B \sim 5 \cdot 10^{15} \text{G}$$

In the proximity of the BH is possible to generate $e^\pm$ pairs than could give origin to the GRB fireball, provided a sufficiently clean environment in order to avoid previous electric field discharge.

# ENERGETICS, TIME DURATION AND VARIABILITY

In the following we assume as hypothesis that the energy source of the GRB is the gravitational collapse of a torus of 0.1 $M_\odot$ onto a rotating BH of 10 $M_\odot$. The available energy from this inelastic collision derived in a semiclassical approximation is $\Delta E \sim 8 \cdot 10^{53}$ erg. The interaction between the torus and the BH is assumed to be the BZ mechanism with a magnetic field of the order of $10^{15}$ G.

The accreted material which releases its gravitational energy gives origin to a variable magnetic field that breaks the vacuum in a volume $\sim R_s^3$. The number of $e^\pm$ pairs produced is:

$$N_{e^\pm} = 2\frac{R_s^3}{\bar{\lambda}^3} = 0.84 \cdot 10^{51}$$

The particle density is $\sim 3 \cdot 10^{31}$ $e^\pm$ cm$^{-3}$. If we assume that the magnetic energy $\frac{B^2}{2\mu_0} \sim 4 \cdot 10^{28}$ erg cm$^{-3}$ is shared among them, each particle gets an energy $\varepsilon_0 \sim 1.3\eta_{\rm acc} \cdot 10^{-3}$ erg, where $\eta_{\rm acc}$ is the acceleration efficiency. Each particle has an energy $\gamma_0 m_e c^2$, so it is possible to derive the relativistic Lorentz factor $\gamma_0 = \frac{\varepsilon_0}{m_e c^2} \sim 1.5\eta_{\rm acc} \cdot 10^3$.

After the formation of the plasmoid containing $\sim 10^{51}$ $e^\pm$ pairs in a volume $V \sim R_s^3$, the particles undergo three important processes:

1) Particle acceleration in a time scale $t_{\rm acc} \sim 10^3 \cdot \frac{\bar{\lambda}}{c}$ to acquire an energy $\sim 10^3 \cdot mc^2$, since each particle acquires on average an energy $\sim mc^2$ every $\bar{\lambda}$.

2) Momenta randomization in a time scale $t_{\rm random} \sim \frac{l}{c}$, where $l$ is the mean free path for $e^\pm$ interaction. $l = \frac{1}{\sigma n} \sim 0.3$ cm, using $\sigma = \frac{87({\rm nb})}{E({\rm GeV})^2}$ and $n = 3.5 \cdot 10^{31}$ cm$^{-3}$.

3) Single particle collimation by synchrotron radiation. The particles momentum components normal to the magnetic field $p_\perp$ are damped in a time scale $t_{\rm coll} < \frac{\rho}{c}$, where $\rho$ is the curvature radius. $\rho \simeq \frac{0.3 E_{\rm GeV}}{c \cdot B_T}$ m. With the presence of a magnetic field of the order of $10^{15}$ G, the particles radiate all the energy corresponding to $p_\perp$ on a small fraction of a turn.

The momentum components perpendicular to the residual field line outside the plasmoid for all the particles are damped in a time $\sim \frac{R_s}{c}$ and the plasmoids becomes a stream of particles with velocity parallel to the external field lines with $\gamma \sim \frac{1}{3}\gamma_0$. As a result the plasmoid travels as a parallel stream with bulk Lorentz factor $\Gamma = \gamma \sim 500\eta_{\rm acc}$.

The energy of the particles in the plasmoid before the cooling by synchrotron emission is:

$$E_{\rm particles} = 2\frac{R_s^3}{\bar{\lambda}^3} \cdot m_e c^2 \Gamma \sim 3.4\eta_{\rm acc} \cdot 10^{47} {\rm erg}$$

The available energy in the overall inelastic collision is $\Delta E \sim 8 \cdot 10^{53}$ erg, so that the emission of plasmoids could happen $N_{\rm plasmoid}$ time where:

$$N_{\rm plasmoid} \sim \eta_B \frac{\Delta E}{E_{\rm plasmoid}} = 2.4\eta_B \cdot 10^6$$

where we have taken into account also the efficiency, $\eta_B$, for conversion of mechanical energy into the magnetic field, and $E_{\text{plasmoid}} = E_{\text{particles}}/\eta_{\text{acc}}$ is the total energy available in each plasmoid.

Since the time duration of the engine is smaller or equal to the observation time $t_{\text{obs}} \sim 10^2$s, the average time separation between two consecutive plasmoids is $t_s \sim t_{\text{obs}}/N_{\text{plasmoid}} \sim 4\eta_B^{-1} \cdot 10^{-5}$s. In the fireball model the prompt $\gamma$ emission of the GRB is due to internal shocks between plasmoids proceeding at different speed.

## CONCLUSIONS

This models predicts for long duration GRB a pulsed emission from $10^5$ emitted plasmoids (with $\eta_B \sim 0.1$) with an average time separation $\Delta t \sim \frac{t_{\text{obs}}}{N_{\text{plasmoids}}} \sim 10^{-4}$ s, corresponding to a distance in space of $\sim 1R_s$. The train of "sausage" plasmoids is $\sim 10^{10}$m at the end of engine activity. Its lenght is slightly modified during the internal shock phase so that the overal time duration is dominated by the duration of engine activity, the shortest time variability instead is determined by the plasmoids interactions.

## ACKNOWLEDGMENTS

We thank A.Celotti, G.Cocconi and M.Tavani for critical reading and valuable comments and G.Fishman and N.Gehrels for the encouragement.

## REFERENCES

1. Mészáros P. (2001), Science, **291**, 79
2. Frail D.A. *et al.* (2001), Nature submitted, `astro-ph/0102282`
3. Amati L. *et al* (2000), Science **290**, 953.
4. Piro L. *et al* (2000), Science **290**, 955.
5. Antonelli L.A. *et al.* (2000), ApJ **545**, L39
6. Rees M.J. & Mészáros P. (2000), ApJ **545**, L73
7. Vietri M. *et al* (2001), ApJ **550**, L43
8. Woosley S.E. (1993), ApJ **405**, 273
9. Paczynski B. (1993), in *Texas/Pascos 92*, Akerlof C.W. and Srednicki M.A. eds., Annals of the New York Academy of Sciences **688**, 321
10. Paczynski B. (1998), ApJ **494**, L95
11. Vietri M. and Stella L. (1998), ApJ **507**, L45
12. Blandford R.D. & Znajek R.L. (1977), MNRAS **179**, 433
13. Thorne K.S., Price R.H. and Macdonald D.A.(1986), *Black holes: the membrane paradigm*, Yale University Press
14. Lee H.K. *et al.* (2000), Physics Reports **325** 83
15. Ruffini R. (1998), `astro-ph/9811232`, in *Black holes and high energy astrophysics*, Sato, H. and Sugiyama, N., eds. Proceedings of the Yamada conference XLIX.